\documentstyle[aps,psfig]{revtex}
\begin{document}
\title{Phenomenological theory of magnetic anisotropy\\
in thin films and multilayers}
\author{A.N.\ Bogdanov$^{1*}$
, U.K. R\"o{\ss}ler$^{2\#}$
}
\address{
$^1$Max-Planck-Institut f{\"u}r Physik komplexer Systeme\\
N\"othnitzer Stra{\ss}e 38, D-01187 Dresden, Germany\\
$^2$Institut f\"ur Festk\"orper-- und Werkstofforschung Dresden\\
Postfach 270016, D--01171 Dresden, Germany}
\date{\today}
\maketitle
\begin{abstract}
Within a phenomenological approach the density of induced uniaxial
anisotropy in nanostructures $K({\bf r})$ is treated as a physical field
additional to the magnetization ${\bf M}({\bf r})$. The equilibrium
distributions of $K({\bf r})$ are formed under the 
influence of the surfaces and internal interactions. 
The functions $K({\bf r})$ and ${\bf M}(%
{\bf r})$ are calculated for a magnetic layer between 
two nonmagnetic
spacers. It is shown that the transition from the 
phase with magnetization in plane to the perpendicular phase 
occurs continuously via 
an intermediate phase which is inhomogeneous 
across the layer-thickness. 
The theory explains experimentally observed
thickness dependences of the effective anisotropy 
and gives a method to obtain values of the characteristic 
parameters for the theory from experimental data. 
It is shown that the separation 
of the induced anisotropy into volume and surface contribution 
is valid only for thick films.
\end{abstract}
%
Magnetic anisotropy is one of the key properties exploited in ferromagnetic
nanostructures. Reduced dimensionality may greatly alter the effective
magnetic anisotropy of such structures. For layered systems, numerous
investigations demonstrate that reduced thickness and complex interactions
on surfaces and interfaces induce uniaxial magnetic anisotropy 
which, by orders of magnitude, exceeds 
the values of the intrinsic anisotropy in the
corresponding bulk materials 
\cite{deJonge94,Johnson96,Poul99}. 
During the last decade a large variety of magnetic thin films and
multilayers have been synthesized and investigated experimentally. They have
already found applications in modern magnetoelectronics such 
as spin-valves or magnetoresistive heads \cite{Himpsel}.

Recent first principle numerical calculations have proved the reliability of
the quantum-mechanical microscopic theory in its applications to magnetic
nanostructures 
\cite{Guo97,Weinberger00}.
They are offering
considerable insight into the phenomena of induced anisotropy. However,
these {\it ab initio} calculations are still unable to give a complete
description of the magnetization structures and processes in real layered
systems. Hence, our understanding of and control 
over the magnetic anisotropy of nanostructures is rather incomplete. 
Up to now, the analysis
of experimental data on anisotropy effects in magnetic layered systems is
mostly based on the effective volume anisotropy method introduced by 
N{\'e}el about fifty years ago \cite{Neel54}. 
Within this approach the value of
average induced uniaxial anisotropy 
in a magnetic layer of thickness $d$ is
given by an empirical ansatz 
\begin{equation}
K_{\mbox{\footnotesize eff}}=K_V+\frac{K_S}d  \label{ansatz}
\end{equation}
where $K_V$ is the volume anisotropy (per unit volume), 
and $K_S$ is the
surface (interface) contribution (per unit area). 
For many layered systems
with reduced thickness, the function $K(1/d$) becomes strongly
nonlinear which is at variance with eq.~(\ref{ansatz}) 
\cite{deJonge94,Johnson96,Broeder91,Jungblut94,Bochi96,Ha00}. 
Moreover, the general validity of the N{\'e}el theory has been questioned
in a number of publications 
\cite{deJonge94,Rado}. 
The separation of
the uniaxial anisotropy into volume and surface contribution
and the reduction of surface
anisotropy into an effective volume contribution 
should be considered as an excessive simplification. 
Furthermore, equation (\ref{ansatz}) implies
that the anisotropy is constant within a magnetic layer and, 
thus, stabilizes a homogeneous distribution of the magnetization. 
It was shown,
however, that the competition between volume and surface anisotropies may
induce inhomogeneous states in magnetic films and multilayer structures \cite
{Thiaville92}. The surface induced anisotropy can also gradually relax 
into the depth of the layer \cite{Kingetsu}. 
On the whole, the heuristic N{\'e}el
approach as well as similar models have 
succeeded to give qualitative
explanations of some general features of 
the induced anisotropy. However,
they do not account for many other observed effects 
and they fail to give a
quantitative description of the magnetization 
structures in magnetic nanostructures.

In this paper we propose 
a simple phenomenological theory which gives a
consistent method to calculate the anisotropy and the equilibrium
magnetization in thin films and multilayers. Let us consider a magnetic
nanostructure embedded into nonmagnetic media. It may be a thin magnetic
film confined by the interface with the substrate and the surface, or a
magnetic layer between nonmagnetic spacers in a multilayer, or any other
magnetic inclusion in a nonmagnetic matrix (such as nanowires, 
magnetic dots, or magnetic clusters). 
The magnetic energy of such a nanostructure can
be written as an integral over its volume 
\begin{equation}
W_m=\int \left[ A\sum_i \left( \frac{\partial {\bf m}}{\partial x_i}\right)
^2+K({\bf r})({\bf m\cdot n})^2-{\bf H\cdot M}
-\frac 12\,{\bf H}_d\cdot {\bf M}
\right] d{\bf r}\,.  \label{func1}
\end{equation}
Here ${\bf m}={\bf M/}M_0$ is the normalized value of the magnetization
vector ${\bf M}$ ($M_0=\left| {\bf M}\right| $), ${\bf n}$ is a unity vector
perpendicular to the surface of the layer, $A$ is the exchange stiffness
constant, ${\bf H}$ is an external magnetic field, and ${\bf H}_d$ is a
demagnetizing field. The second term in (\ref{func1}) is the energy density
of the induced uniaxial anisotropy. To describe the distribution of $K$($%
{\bf r}$) within the nanostructure we introduce an interaction functional of
Landau-Ginzburg type 
\begin{equation}
W_A=\int \left[ {\it \alpha }\sum_i \left( \frac{\partial K}{\partial x_i}%
\right) ^2+f(K)\right] d{\bf r}\,. \label{func2}
\end{equation}
The first term in (\ref{func1}) represents the stiffness energy and $f$($K$)
is the energy of a homogeneous induced anisotropy. In many cases the
function $f$($K$) may be written as $f(K)=aK^2+2bK$.
The stiffness parameter ${\it \alpha }>0$ together with $a>0$ characterize
the resistance of the system against the uniaxial anisotropy imposed by the
boundaries. The linear term in $f$($K$) reflects the tendency for symmetry
breaking and the ensuing rise of homogeneous uniaxial anisotropy.
Minimization of $f$($K$) yields the value of this anisotropy $K_b=-b/a$ .
The scalar function $K({\bf r})$ may be treated as a physical field
additional to the magnetization field ${\bf M}({\bf r})$. Both fields are
coupled by the anisotropy energy in (\ref{func1}). In most cases, this
coupling energy should be negligible compared to the strong surface and
volume interactions inducing the uniaxial anisotropy. Thus, the equilibrium
distribution $K$(${\bf r}$) can be calculated independently from the
magnetization field by minimization of the functional (\ref{func2}). The
anisotropy on the confining surfaces is supposed 
to have values $K({\bf r})\!\!\!\mid_S\,=K_0$
fixing the boundary conditions for the
variation problem. 
Then, the equilibrium distributions of the magnetization
are determined by variation of the functional (\ref{func1}) 
with this
definite function $K$(${\bf r}$). 
Here, the boundary conditions $K_0$ are kept constant.
But, in general, $K_0$ may vary within the
surfaces, and all phenomenological constants
in (\ref{func2}) may depend on characteristic 
sizes of the nanostructure. 
The values $K_0$ (volume anisotropy density on the surfaces) 
should not be confused with the ``surface anisotropy'' $K_S$ 
in (\ref{ansatz}) which is an
anisotropy energy per area. The relation between these quantities
will be discussed below.

As an application of the theory we calculate the induced anisotropy and the
magnetization in a magnetic layer of thickness $d$ sandwiched between two
identical nonmagnetic spacers. The layer is supposed to be infinite in $x$
and $y$ directions and bounded by parallel planar surfaces $z=\pm d/2$. On
the surfaces the anisotropy is $K(d/2)=K(-d/2)=K_0$. The vector ${\bf n}$ is
directed along the $z$-axis and $K_0>0$ (perpendicular, surface induced
anisotropy). Within the magnetic plate $K$(${\bf r}$) varies along $z$%
-direction and is homogeneous in ($x,y$) plane. Variation of the functional (%
\ref{func2}) yields the following solution for $K$($z$): 
\begin{equation}
K(z)=K_b+\frac{\left( K_0-K_b\right) \cosh \left( z/{\it \lambda }_a\right) 
}{\cosh \left( d/(2{\it \lambda }_a)\right) }  \label{Kz}
\end{equation}
where ${\it \lambda }_a=\sqrt{{\it \alpha }/a}$ is a characteristic length.
To derive the equilibrium functions ${\bf M}({\bf r}$) 
one has to minimize
the functional for the magnetic energy (\ref{func1}) 
with $K$($z$) given by (\ref{Kz}). 
We consider structures that are homogeneous in the plate plane
(i.e. monodomain structures). Further, we ignore the variation of $M_0$
observed in some multilayers \cite{Wilhelm}. 
Under these assumptions the magnetization in the layer 
is described by the polar angle ${\it \theta }(z)$
between the vector ${\bf M}$ and $z$-axis. For such a one-dimensional
problem the magnetostatic equations for ${\bf H}_d$ have a rigorous
solution, and the corresponding stray field energy is expressed by an
anisotropy-like term 
$w_d=-{\bf H}_d{\bf M}/2=2\pi M_0^2\cos ^2({\it \theta })$
\cite{Hubert98}. This energy contribution is commonly named ``shape
anisotropy''.
Thus, the functional (\ref{func1}) with $K(z)$ given by (\ref{Kz}) may be
expressed in the following reduced form 
\begin{equation}
W=2\pi M_0^2\int_{-d/2}^{d/2}\left\{ {\it \lambda }_{ex}^2\left( \frac{d{\it %
\theta }}{dz}\right) ^2-\frac{\widetilde{K}(z)}{2\pi M_0^2}\cos ^2{\it %
\theta }-h\cos ({\it \theta -\psi })\right\} dz  \label{func3}
\end{equation}
where $h=H/(2\pi M_0)$ is a reduced value of the magnetic field and ${\it %
\psi }$ is the angle between the vector ${\bf H}$ and $z$-axis, ${\it %
\lambda }_{ex}=\sqrt{A/\left( 2\pi M_0^2\right) }$ is the exchange length,
and the volume density of the total anisotropy $\widetilde{K}(z)$ is 
\begin{equation}
\widetilde{K}(z)=-2\pi M_0^2+\frac{K_0\cosh \left( z/{\it \lambda }_a\right) 
}{\cosh \left( d/(2{\it \lambda }_a)\right) }\,.  \label{anisotropy}
\end{equation}
For simplicity we suppose that the homogeneous part of the induced
anisotropy $K_b$ (see (\ref{Kz})) is included into the coefficients $K_0$
and $2\pi M_0^2$.

The energy (\ref{func3}) has 
the functional form of an uniaxial ferromagnet
with anisotropy $\widetilde{K}(z)$ (\ref{anisotropy}) 
varying along the layer thickness. 
The distribution of $\widetilde{K}(z)$ within the layer
strongly depends on the layer thickness (Fig. 1). 
The length ${\it \lambda }_a$ 
characterizes the extension of the surface anisotropy into the layer
depth and may be named {\it ``penetration depth of the induced
anisotropy''}. When $d\gg{\it \lambda }_a$ the induced anisotropy is
localized near surfaces and rapidly relaxes into the depth of the layer. 
For thinner layers, where ${\it \lambda }_a$ and $d$ 
have the same order of magnitude, the induced anisotropy 
attains considerable values within all the
layer (Fig. 1).
The possible magnetic phases and the regions of their stability are
determined by optimization of the energy (\ref{func3}) with free boundary
conditions that read for this case $\left( d{\it \theta }/dz\right) \mid
_{z=\pm d/2}=0.$ A complete analysis of the functional (\ref{func3}) will be
done elsewhere. Here we only discuss some general properties of the
solutions at zero field as representative for the problem. In this case the
layer thickness $d$ and the parameters $q=K_0/\left( 2\pi M_0^2\right) $ and 
${\it \Lambda }={\it \lambda }_{ex}/{\it \lambda }_a$ span the phase space
of the solutions. A typical ($q$, $d$) phase diagram for fixed values of $%
{\it \Lambda }$ is shown in Fig.~2. It consists of two regions where
homogeneous phases exist with either 
{\em perpendicular} (${\it \theta }=0$)
or {\it parallel } (${\it \theta }=\pi /2$) magnetization. 
These are separated by an inhomogeneous ({\em twisted}) phase 
where the angle ${\it \theta }$ gradually increases 
from a definite value ${\it \theta }_1\geq 0$
on the surfaces to a largest value in 
the center ${\it \theta }_1<{\it \theta }_2\leq \pi /2$.

The perpendicular phase is stable for small values of the thickness and when 
$q>1$. The parallel phase exists only in rather thick layers and when the
parameter $q$ is smaller than a critical value $q_c({\it \Lambda })$. In the
region $q>q_c$ the perpendicular phase transforms continuously into the
twisted phase at the critical line $d_1(q)$. For $1<q<q_c$ the region of the
existence of the twisted phase is bound by two lines of second order
transitions into the homogeneous phases $d_1(q)$ and $d_2(q)$. Thus,
according to our model the transition between the parallel and the
perpendicular phases always occurs continuously via the transformation into
an intermediary (twisted) phase. Fig.~3 shows the evolution of the magnetic
structures during such a transition.

By integrating $\widetilde{K}(z)$ in (\ref{anisotropy}) over $z$ one obtains
for the average value of the uniaxial anisotropy density 
\begin{equation}
K_{\mbox{\footnotesize eff}}=\frac 1d\int_{-d/2}^{d/2}\widetilde{K}%
(z)dz=-2\pi M_0^2+\frac{K_0\tanh \left( d/(2{\it \lambda }_a)\right) }{d/(2%
{\it \lambda }_a)}\,.  \label{Keff}
\end{equation}
Values of the anisotropy $K_{\mbox{\footnotesize eff}}$ 
have been measured in many layered systems 
\cite{deJonge94,Johnson96}. 
Usually, experimental values of $K_{\mbox{\footnotesize eff}}$ 
are plotted as product $K_{\mbox{\footnotesize eff}}\,d$ versus $d$. 
In Fig.~4 the thickness dependence of 
the function ${\it \Phi }(d)=K_{\mbox{\footnotesize
eff}}\, d$ corresponding to (\ref{Keff})
is shown for different values $q$.

In the limit of large thickness ($d\gg{\it \lambda }_a)$ 
the functions ${\it \Phi }(d)$ are negative 
and depend linearly on the thickness (with
slope equal to the value of the shape anisotropy). 
In thiner layers ($d\leq {\it \lambda }_a$) 
the functions ${\it \Phi }(d)$ become nonlinear. 
For $q<1$ they are always negative (parallel anisotropy) 
and reach zero as the thickness decreases to zero. 
For $q>1$, the function ${\it \Phi }(d)$
changes sign at the critical thickness $d_0$ 
determined by the equation 
\begin{equation}
K_0\tanh \left( \frac{d_0}{2{\it \lambda }_a}\right) =\frac{2\pi M_0^2d_0}{2%
{\it \lambda }_a}  \label{eq1}
\end{equation}
and remains positive for $0<d<d_0$ manifesting the existence of the
perpendicular anisotropy in this region. The function ${\it \Phi }(d)$
reaches the maximum value for the thickness 
\begin{equation}
d_{\max }=2{\it \lambda }_a\,
\mbox{arc}\cosh \sqrt{\frac{K_0}{2\pi M_0^2},}
\label{eq2}
\end{equation}
and then monotonically decreases 
to zero as $d\rightarrow 0$ (Fig.~4).
Experimentally obtained functions ${\it \Phi }(d)$ 
are in close accordance
with our theoretical results. 
E.g., nonmonotonic dependencies of ${\it \Phi }(d)$ 
have been observed for Co/Au multilayers\cite{Kingetsu}, 
Co/Ir multilayers\cite{Broeder91}, 
or for Ni/Cu multilayers 
and Cu/Ni/Cu sandwiches
\cite{Jungblut94,Bochi96,Ha00}. 
Taking experimental values for $d_0$
and $d_{\max }$ it is possible 
to calculate the parameters $q$ and ${\it \lambda }_a$ 
from eqs.~(\ref{eq1}) and (\ref{eq2}). 
E.g., for Co/Au \cite{Kingetsu} films 
this yields $q=4.6$ and ${\it \lambda }_a=1.9$\AA . 
For Ni/Cu multilayer systems 
the function${\it \Phi }(d)$ 
in the regions of perpendicular anisotropy 
has been obtained by three different 
experimental methods 
\cite{Jungblut94,Bochi96,Ha00}. 
The results from these experiments
show similar functional dependencies 
for ${\it \Phi }(d)$.
However, the measured values display considerable quantitative 
differences. Taking the experimental data from \cite{Jungblut94}
and \cite{Bochi96} we derive the following values for 
the characteristic parameters $q=2.1$ 
and $2.2$, ${\it \lambda }_a=26.4$\AA\ and $31.9$\AA . 
In contrast to the full curves from our theory 
in Fig.~4 experimental dependencies ${\it \Phi }(d)$ 
for these Ni-films become negative
again for finite values of the thickness manifesting 
the reentrance of planar anisotropy 
in ultrathin films \cite{Jungblut94,Bochi96,Ha00}. 
It should be noted in this respect that 
for the calculation of ${\it \Phi }(d)$ (Fig.~4) 
the phenomenological parameters $K_0$ are kept constant
in this simple approach. According to experimental observations 
they may depend on the layer thickness, and they may be sensitive 
to details of the film preparation, 
e.g. via surface reconstructions, misfit strain 
or misfit dislocations etc.
In particular, it was established experimentally 
that in the Ni-films under discussion surface anisotropy 
decreases with decrease of the layer thickness
due to magnetoelastic interactions \cite{Bochi95}. 
For decreasing values of $K_0$ the theoretical 
dependencies ${\it \Phi }(d)$ become negative in the
limit of very thin films (dashed line in Fig.~4). 
This thickness dependency
of the phenomenological parameters $K_0$ 
as well as other phenomenological
constants should be considered 
by introducing appropriate models 
in future work based on further experimental results 
and results from {\em ab initio} calculations.

The distribution of the induced anisotropy 
and the magnetization in a magnetic layer can also 
be investigated within a discretized model
which is useful in the limit of ultrathin layers 
or for comparison with {\em ab initio} calculations. 
In this case the equilibrium values of the induced anisotropy 
in the magnetic layer consisting of $N$ magnetic planes 
are determined by minimization of the energy 
\begin{equation}
\widetilde{W}_A=\sum_{n=1}^{N-1}\left[ \left( 1+\frac{2{\it \lambda }_a^2}{%
\Delta ^2}\right) K_n^2-\frac{2{\it \lambda }_a^2}{\Delta ^2}K_nK_{n+1}+%
\frac{2b}aK_n\right] \Delta  \label{discrete}
\end{equation}
where $\Delta $ is the distance between two adjacent planes. Equilibrium
profiles of $K_n$ for symmetric boundary conditions $K_1=K_N=K_0$ are shown
in Fig.~5. Using these profiles, the orientation of the magnetization
vectors in the planes ${\it \theta }_n$ can be calculated by minimization of
the discretized analogue of the energy (\ref{func3}). 
The characteristic functions for the discretized model
\(
{\it \Phi }_N=-2\pi M_0^2N\Delta +\sum_{n=1}^NK_n 
\)
shown in Fig.~6 display the same dependence on thickness 
$d=N\Delta $ as the continuous model (Fig.~4).

Finally, we discuss the relation 
between our theory and the traditional approach.
For large thickness ($d\gg {\it \lambda }_a)$, 
the function (\ref{Kz}) may be asymptotically written 
as 
\begin{equation}
K(z)=K_0\exp (-\frac{d-\left| z\right| }{{\it \lambda }_a})\,.
\label{asympt}
\end{equation}
The function $K(z)$ has finite values only 
in the vicinity of the surfaces and equals zero 
in the main part of the layer. Integrating (\ref{asympt})
with respect to $z$ yields the anisotropy energy (per area) 
\begin{equation}
K_s=K_0\int_0^\infty \exp (-{\it \xi }/{\it \lambda }_a)\; d{\it \xi}%
=K_0 \, {\it \lambda }_a \,.  \label{Ks}
\end{equation}
The integral parameter $K_s$ is equal to the additional 
energy connected with surface induced anisotropy. 
Similar parameters characterize surface
tension in liquids and domain wall energy 
in the theory of magnetic domains.
Expanding $K_{\mbox{\footnotesize eff}}$ 
in (\ref{Keff}) for large $d/{\it \lambda }_a$ 
and substituting for $K_s$\ (\ref{Ks}) one obtains the N{\'e}el
anisotropy ansatz (\ref{ansatz}). In the region of large thickness the
surface anisotropy may be transformed 
into a surface term which should be included 
into the boundary conditions of 
the corresponding variation problem
for the magnetization. 
For this case the calculation of the magnetization 
in magnetic layers firstly has been done by Thiaville and Fert 
\cite{Thiaville92}.
It is clear that the surface anisotropy $K_s$ 
is meaningful only in the case of a strong confinement 
of the induced anisotropy to the surface, 
i.e. when $d\gg{\it \lambda }_a$. 
This relation determines the region of the validity
for the N{\'e}el approach and other approaches 
that reduce surface effects
to an effective surface energy contribution. 
Beyond this region the function 
$\widetilde{K}(z)$ of eq.~(\ref{anisotropy}) 
has finite values within all the layer
and the separation of the induced anisotropy 
into a surface and a volume
contribution becomes meaningless.

In conclusion, our theory considers inhomogeneous 
distributions of induced
magnetic anisotropy. It is able to describe many of 
the experimental features found for thin ferromagnetic films. 
In particular, the nonlinear
deviations from N{\'e}el's approach for surface anisotropy found
in various experiments are consistently explained. 

We thank H.\ Eschrig, R.\ Hayn, K.-H.\ M{\"u}ller,  
and P.M.\ Oppeneer for discussion.  
A.N.B.\ thanks P.\ Fulde for hospitality 
and support. U.K.R.\ is supported by DFG.

\clearpage
\begin{figure}[tbp]
\caption{Inhomogeneous distribution 
of the anisotropy $\widetilde{K}(z)$
across a magnetic film according to eq.~(6). 
For the curves from the top to
the bottom, the chosen value of $\lambda_a$ decreases. }
\end{figure}
\begin{figure}[tbp]
\caption{Typical phase diagram (schematic) for the magnetization
structure of magnetic films described by eqs.~(5) and (6) 
for fixed $\Lambda=\lambda_{ex}/\lambda_a$ and in zero field.
Depending on layer thickness $d$ and on
the strength of induced uniaxial anisotropy at the film
boundaries $q \sim K_0$ three phases may exist as symbolically
depicted by the insets. A twisted phase is separated
from the two homogeneous phases by
second-order transition lines $d_1$ and $d_2$.
At the dashed line
the energy for perpendicular and parallel phase
would be equal.
For $q > q_c$ only perpendicular and twisted phase exist.}
\end{figure}
\begin{figure}[tbp]
\caption{ Magnetization distribution across magnetic films for fixed $q$
with increasing thickness $d$. Transition from perpendicular to twisted
structure at $d_1$ and from twisted to parallel at $d_2$. }
\end{figure}
\begin{figure}[tbp]
\caption{
Characteristic function
$\Phi(d)=K_{\mbox{\scriptsize eff}}\,d$
showing the dependence of
the effective magnetic anisotropy eq.~(7)
on film thickness.
Full lines give the typical behaviour
when the anisotropy $K_0$ at the
boundaries is fixed.
The dashed line shows reentrance of the parallel
magnetization when $K_0$ decreases
with decreasing $d$ in the case with $q>1$. }
\end{figure}
\begin{figure}[tbp]
\caption{ Distribution of local anisotropy for the discretized model
eq.~(10) with various $N$ and fixed thickness.}
\end{figure}
\begin{figure}[tbp]
\caption{ Examples of characteristic functions
for the effective anisotropy of the discretized model.
Symmetric boundary conditions $K_0$=0.05 ($\Box$),
$K_0$=1 ($\bullet$) and
$\lambda_a^2=10$, $b=0.5$, $a=\Delta=1$, $M_0=0.2$.
}
\end{figure}

\clearpage
\hspace{1.0cm}

\vspace{2.5cm}

\psfig{figure=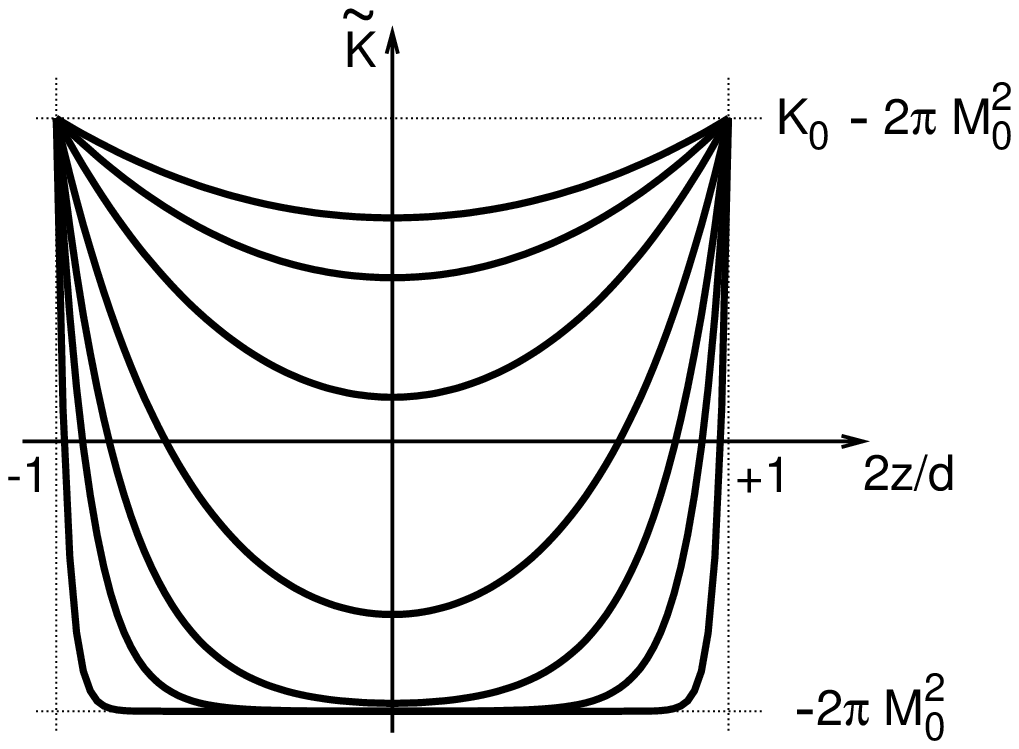,width=10.5cm}

\vspace{4.0cm} {\sloppy{\Large {\bf Fig.~1\\A.N.\ Bogdanov\\Phenomenological
theory of magnetic anisotropy $\dots$ }}}

\clearpage
\hspace{1.0cm}

\vspace{2.5cm}

\psfig{figure=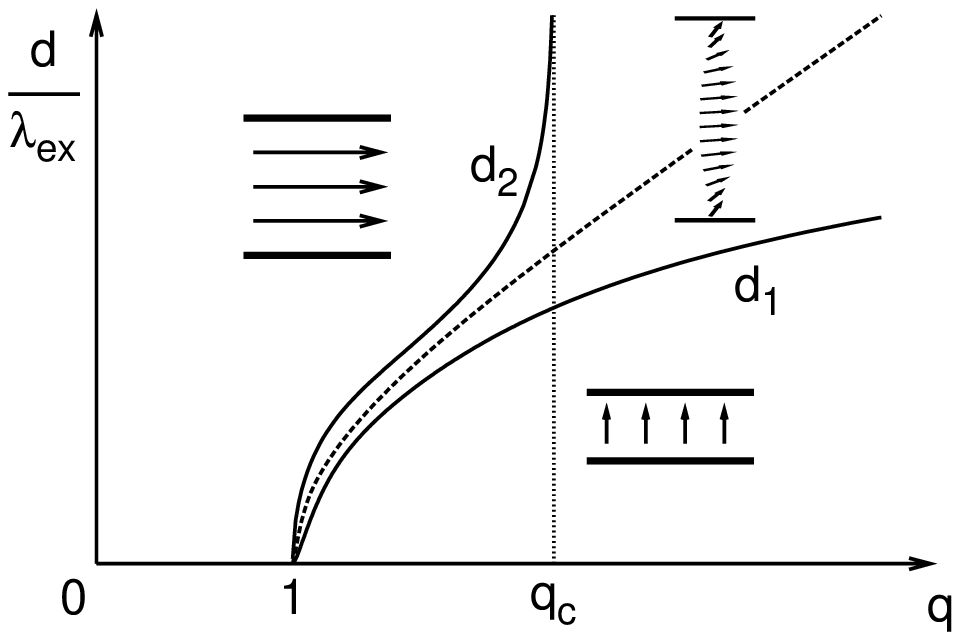,width=10.5cm}

\vspace{4.0cm} {\sloppy{\Large {\bf Fig.~2\\A.N.\ Bogdanov\\Phenomenological
theory of magnetic anisotropy $\dots$ }}}

\clearpage
\hspace{1.0cm}

\vspace{2.5cm}

\psfig{figure=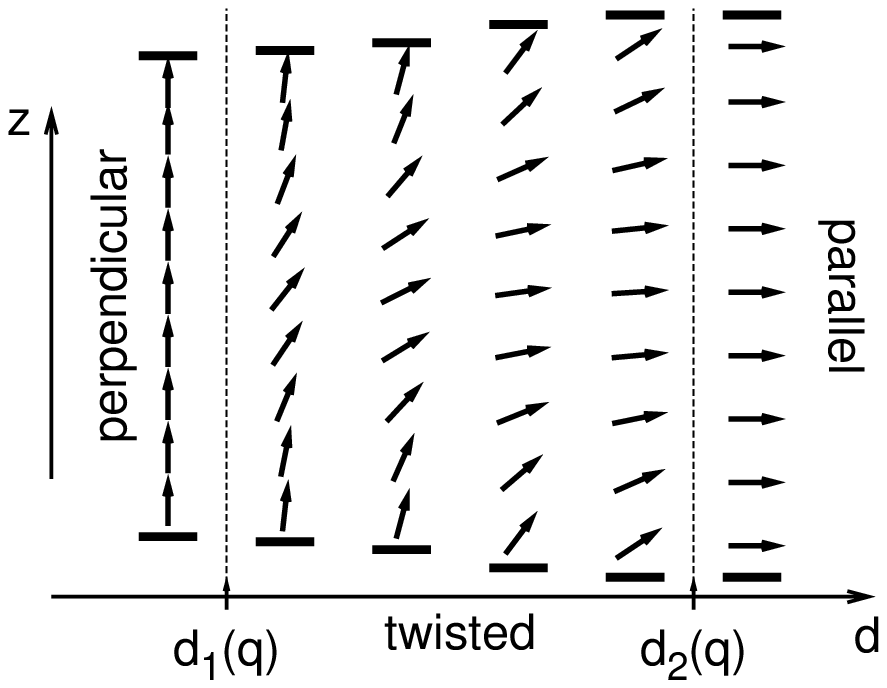,width=10.5cm}

\vspace{4.0cm} {\sloppy{\Large {\bf Fig.~3\\A.N.\ Bogdanov\\Phenomenological
theory of magnetic anisotropy $\dots$ }}}

\clearpage
\hspace{1.0cm}

\vspace{2.5cm}

\psfig{figure=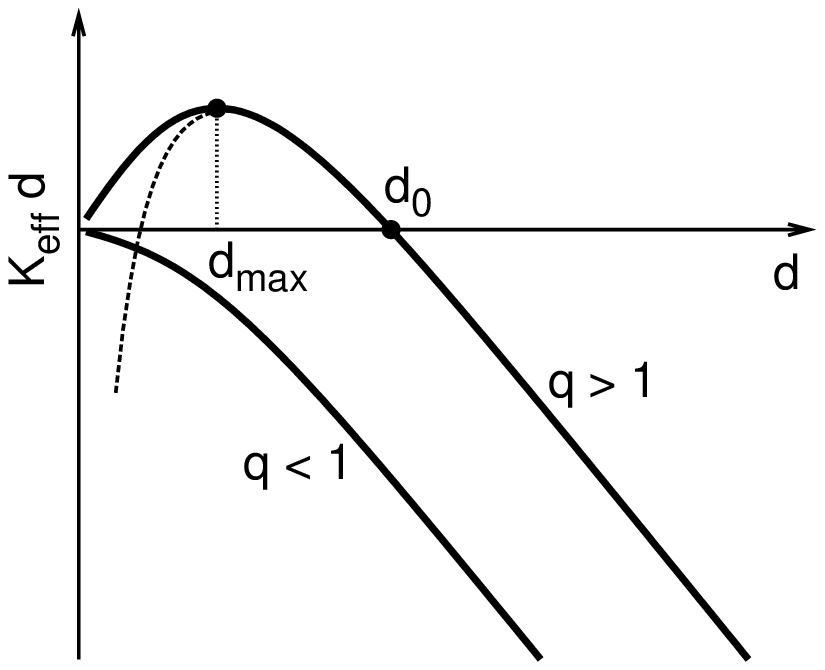,width=10.5cm}

\vspace{4.0cm} {\sloppy{\Large {\bf Fig.~4\\A.N.\ Bogdanov\\Phenomenological
theory of magnetic anisotropy $\dots$ }}}

\clearpage
\hspace{1.0cm}

\vspace{2.5cm}

\psfig{figure=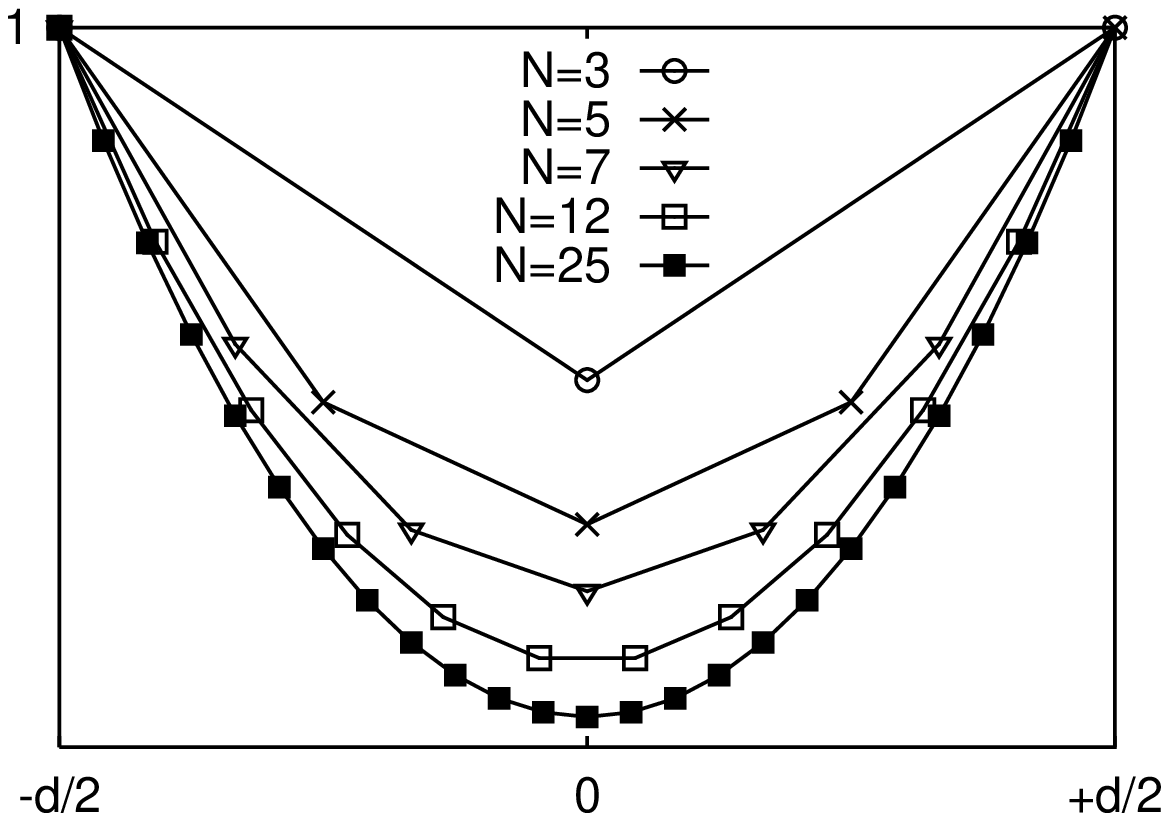,width=10.5cm}

\vspace{4.0cm} {\sloppy{\Large {\bf Fig.~5\\A.N.\ Bogdanov\\Phenomenological
theory of magnetic anisotropy $\dots$ }}}

\clearpage
\hspace{1.0cm}

\vspace{2.5cm}

\psfig{figure=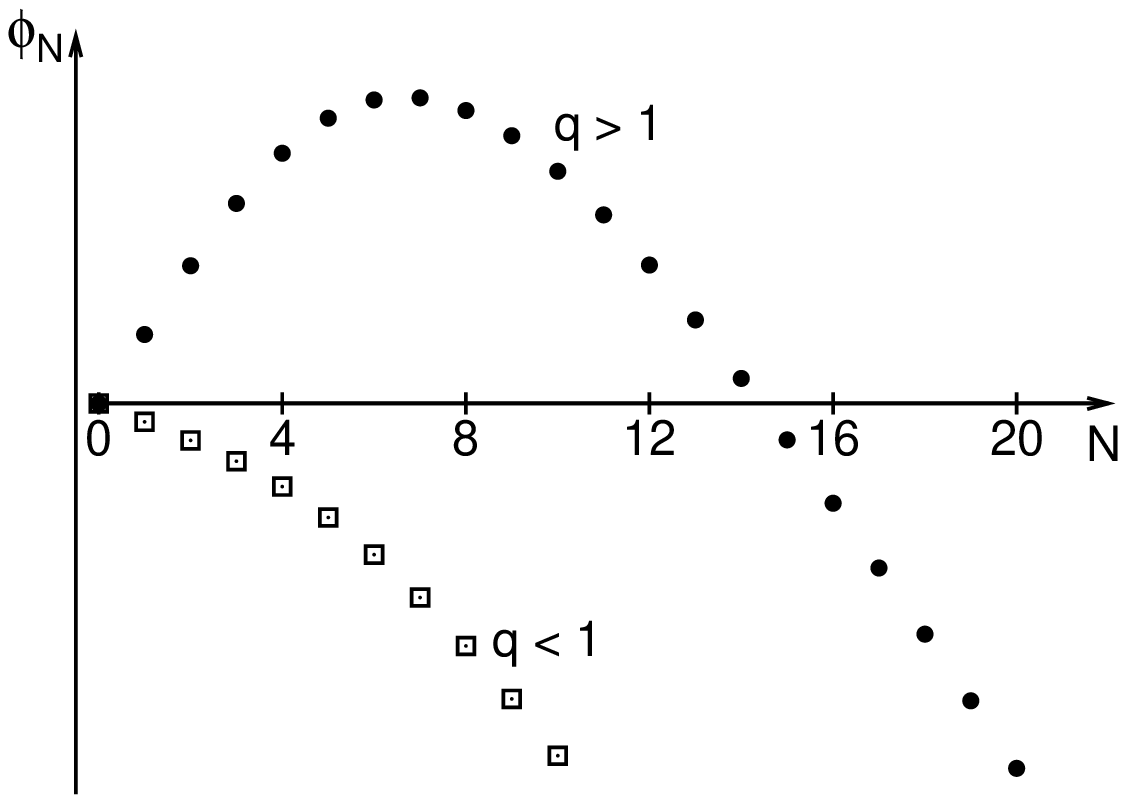,width=10.5cm}

\vspace{4.0cm} {\sloppy{\Large {\bf Fig.~6\\A.N.\ Bogdanov\\Phenomenological
theory of magnetic anisotropy $\dots$ }}}

\end{document}